\documentclass[12pt]{article}

\usepackage{arxiv}

\usepackage{amsmath,amsfonts,amssymb}
\usepackage{graphicx}
\usepackage{setspace}
\usepackage{tocloft}

\usepackage{tikz-timing}

\usepackage{nccmath}
\usepackage{color}
\usepackage{xspace}
\usepackage{amsmath,amssymb}
\usepackage{bm}
\usepackage{graphicx}
\usepackage{cite}
\usepackage{lmodern}
\usepackage{enumitem}
\usepackage{multirow}
\usepackage{array,booktabs}
\usepackage{calligra}
\usepackage{tablefootnote}
\usepackage{footnote}
\usepackage{makecell}
\usepackage{siunitx}
\usepackage{pgfplots}
\usepackage{subfig}
\usepackage{stackengine}
\usepackage{minitoc}
\usepackage{gensymb}
\usepackage{hyperref}  
\usepackage{breakurl}
\usepackage{algorithm}
\usepackage[noend]{algpseudocode}
\pgfplotsset{compat=newest}
\pgfkeys{/pgf/number format/.cd,fixed,precision=3}

\usepackage{xparse}
\usepackage{tikz}
\newcommand{\enquote}[1]{``#1''}

\newcommand{\abs}[1]{\vert #1 \vert}

\newcommand{\map}[3]{#1\colon #2 \to #3}
\newcommand{\Int}[3]{\int_{#1}{#2}\,\mathit d{#3}}
\newcommand{\R}{\mathbb R}

\newcommand{\x}{\bm{x}}
\newcommand{\y}{\bm{y}}

\newcommand{\Per}{\mathrm{Per}}
\newcommand{\img}{\bm{f}}
\newcommand{\edge}{\mathcal{E}} 
\newcommand{\scrib}{\mathcal{S}} 

\usepackage{pifont}

\makeatletter
\renewcommand{\normalsize}{\@setfontsize\normalsize{12}{14}}
\makeatother

\normalsize

\title{Robustness of Brain Tumor Segmentation}

\author{
Sabine M\"uller$^{1,2,3,4}$, Joachim Weickert$^{3}$, Norbert Graf$^{4}$ \\[5mm]
$^{1}$Competence Center High Performance Computing, \\Fraunhofer ITWM, \\
67663 Kaiserslautern, Germany\\
$^{2}$Fraunhofer Center Machine Learning, Germany\\
$^{3}$Mathematical Image Analysis Group\\
Faculty of Mathematics and Computer Science, Campus E1.7\\
Saarland University \\
66041 Saarbr\"ucken, Germany\\
$^{4}$Department of Pediatric Oncology and Hematology\\
Saarland University Medical Center\\
66421 Homburg, Germany 
}

\begin{document}
\maketitle

\begin{abstract}
\noindent
\textbf{Purpose:} The segmentation of brain tumors is one of the most active 
areas of medical image analysis. While current methods perform superhuman on 
benchmark data sets, their applicability in daily clinical practice has not 
been evaluated. In our work we investigate the generalization behavior of deep 
neural networks in this scenario.
\newline\noindent
\textbf{Approach:} We evaluate the performance of three state-of-the-art 
methods, a basic U-net architecture and a cascadic Mumford-Shah approach. 
We also propose two simple modifications (which do not change the topology) to 
improve generalization performance. 
\newline\noindent
\textbf{Results:} In our experiments we show that a well-trained U-network 
shows the best generalization behavior and is sufficient to solve this 
segmentation problem. We illustrate why extensions of this model in a realistic 
scenario can be not only pointless but even harmful.
\newline\noindent
\textbf{Conclusions:} We conclude from our experiments that the generalization 
performance of deep neural networks is severely limited in medical image 
analysis especially in the area of brain tumor segmentation. In our 
opinion, current topologies are optimized for the actual benchmark data 
set, but are not directly applicable in daily clinical practice.
\end{abstract}

\keywords{deep learning, segmentation, generalization, brain tumors}

\section{Introduction}

Since AlexNet \cite{krizhevsky2012imagenet} won the \enquote{ImageNet 
Large   Scale Visual Recognition Competition} challenge 
\cite{deng2009imagenet}, the 
influence of deep neural networks has increased dramatically in all 
do\-mains of image processing and pattern recognition: From 
classification to object tracking and image and video segmentation, 
new approaches are typically based on deep learning strategies 
\cite{HuangLP0HW17,cciccek20163d,bakas2018identifying}. These 
approaches are also gaining more and more influence in the field of 
medical image processing. Since Ronneberger et al. proposed the U-Net 
structure \cite{RFB15a}, this model is the de facto standard method in the 
field of medical image segmentation. While the original approach could be 
trained with relatively few examples in a short time, current models require 
a large amount of data and a very time consuming and computationally 
intense 
trai\-ning cycle \cite{myronenko20183d,isensee2018nnu}. Since several years it 
is common practice to compare the performance of segmentation approaches on 
benchmark data sets. One of the best known data sets has been provided in 
the scope of the \enquote{Multimodal Brain Tumor Segmentation Challenge} 
(BraTS)\index{BraTS} \cite{menze2014multimodal}. \\
Brain tumors account only for a very small fraction of all types of 
cancer, but are also among the most fatal forms of this deadly disease. 
Gliomas, developing from the glial cells, are the most frequent primary brain 
tumors. The fast growing and more aggressive types of gliomas called high-grade 
gliomas, come with an median overall survival rate of up to 15 months 
\cite{marsh2013current}. The standard diagnosis technique for brain 
tumor is magnetic resonance imaging (MRI) \cite{wen2010updated} providing 
detailed information about the tumor and the surrounding brain. Tumor 
segmentation is of crucial importance in surgical and treatment planning, while 
fully automated segmentation is a challenging task especially for high-grade 
gliomas: they usually show diffuse and irregular boundaries and have 
intensities overlapping with normal brain tissue caused by peritumoral edema. 
Moreover, acquisition parameters are not standardized, and different parameter 
settings can have a substantial impact on the visual appearance of the tumor. 
This makes it difficult to compare the quality of different methods for brain 
tumor segmentation. As a step towards an unbiased performance evaluation, the 
BraTS database has been created 
\cite{menze2014multimodal,bakas2017advancing,bakas2018identifying}, and many 
recent approaches report benchmark results on either the full data set or parts 
of it; see e.g. \cite{urban_14_multi-modal,myronenko20183d,isensee2018nnu} . \\
Since this data set has been used for seven years now to compare 
different approaches with each other, a major drawback has manifested itself 
over this long time: The main focus of the researchers is not to present the 
most robust network with best generalization behavior, but to maximize the 
performance metrics of the BraTS benchmark dataset. Thus, it can happen that
the increasingly complicated models are not useful in a real clinical 
scenario as they heavily overfit the test set: this benchmark data set is 
saturated.\\
Nowadays models do not get better in a general sense 
but current best approaches overfit the test set more than others. 
Typically, a test set is meant to be a biased version of a specific 
problem representation, i.e. all humans with high grade brain tumors in MRI 
sequences. In order to show a statistically significance of one benchmark 
result 
being superior to another one, an appropriate sample size is necessary 
\cite{chow2017sample}. Unfortunately, the sample size of the BraTS test 
set is too small to provide a statistical significant difference of the 
best performing methods \cite{bakas2018identifying}.\\
In addition, the main strength of deep learning approaches of fitting 
the underlying data distribution is also their greatest weakness: in a 
clinical setting, the assumption that training and test data belong to the 
exact same distribution is typically not correct. \\
In this work, we evaluate the robustness of different segmentation 
methods with respect to disturbances in the underlying distribution. We 
investigate three state-of-the-art methods as well as a simple and intuitive 
scheme based on the powerful Mumford-Shah functional.\\ 
In addition, we suggest two simple and straight forward modifications 
that allow to increase the generalization performance of the evaluated deep 
neural networks. Finally, we demonstrate that our semi-supervised segmentation 
approach is a powerful post-processing step, that allows to robustify 
the predictions of deep neural networks with respect to disturbances in the 
test data set. Hence, we combine the best of two worlds: We still can learn 
the class distributions of the targeted objects while exploiting the robustness 
of energy formulations to modifications in the data.\\

\subsection{Contributions}
Our contributions are as follows: First, we show that current 
state-of-the-art neural network architectures for
brain tumor segmentation have a poor generalization behavior and massively overfit the training data. Second, we apply two simple but powerful modifications from the classification community to semantic image segmentation. These alterations 
allow for higher generalization performance at inference while 
reducing the network size. Last but not least, we suggest an effective 
post-processing step that massively improves the segmentation result when
disturbances in the data are an issue.

\subsection{Structure of the Paper}
In Section 2, we start with the explanation of different state-of-the-art deep 
neural networks for brain tumor segmentation and a classical approach that does 
not require training and is based on the well-known Mumford-Shah functional. We 
then illustrate two new approaches that help to improve the generalization 
performance of deep neural networks. \\
In Sec. 3 we analyze the sensitivity of the previously presented deep 
neural networks to slight changes in the data distributions. We complete this 
work with our conclusions in Sec. 4.

\section{Materials and Methods}
The baseline of our evaluation is a segmentation approach that does not 
require training: a cascadic Mumford-Shah cartoon model 
\cite{muller2016robust}. Since we can almost eliminate an overfit to the 
underlying data set, none of the compared deep learning models should score 
below the performance of this method.\\ 
We begin our evaluation with the de-facto standard model for image 
segmentation with deep neural networks: the U-Net architecture 
\cite{RFB15a}. We continue with an extended version, namely the No NewNet 
approach showing high performance on BraTS 2018 \cite{isensee2018nnu}. 
Afterwards, we take the winner of last years' challenge into account: NVDLMED, 
using autoencoder regularization to improve the segmentation accuracy 
\cite{myronenko20183d}. Last but not least, we investigate the third place of
the BraTS~2018 challenge, using a cascade of several neural networks for segmentation 
\cite{zhou2018learning}.\\

\subsection{Segmentation Approaches}
\subsubsection{Cascadic Mumford-Shah Cartoon Model}
\label{sec:cms}
The segmentation approach based on a cascadic Mumford-Shah cartoon 
model \cite{muller2016automatic}, does not require training.
Its basic idea is to exploit a simple but reliable prior information: 
Brain tumors are usually brighter than the surrounding brain tissues
on $T_2$-Flair images. Typically, it is difficult to identify a good
parameter setting for these kind of segmentation approaches. Fortunately,
our method only depends on a single parameter choice, that can be 
automatically selected. 

\begin{figure}[t]
  
  \begin{center}
\includegraphics[height=6cm]{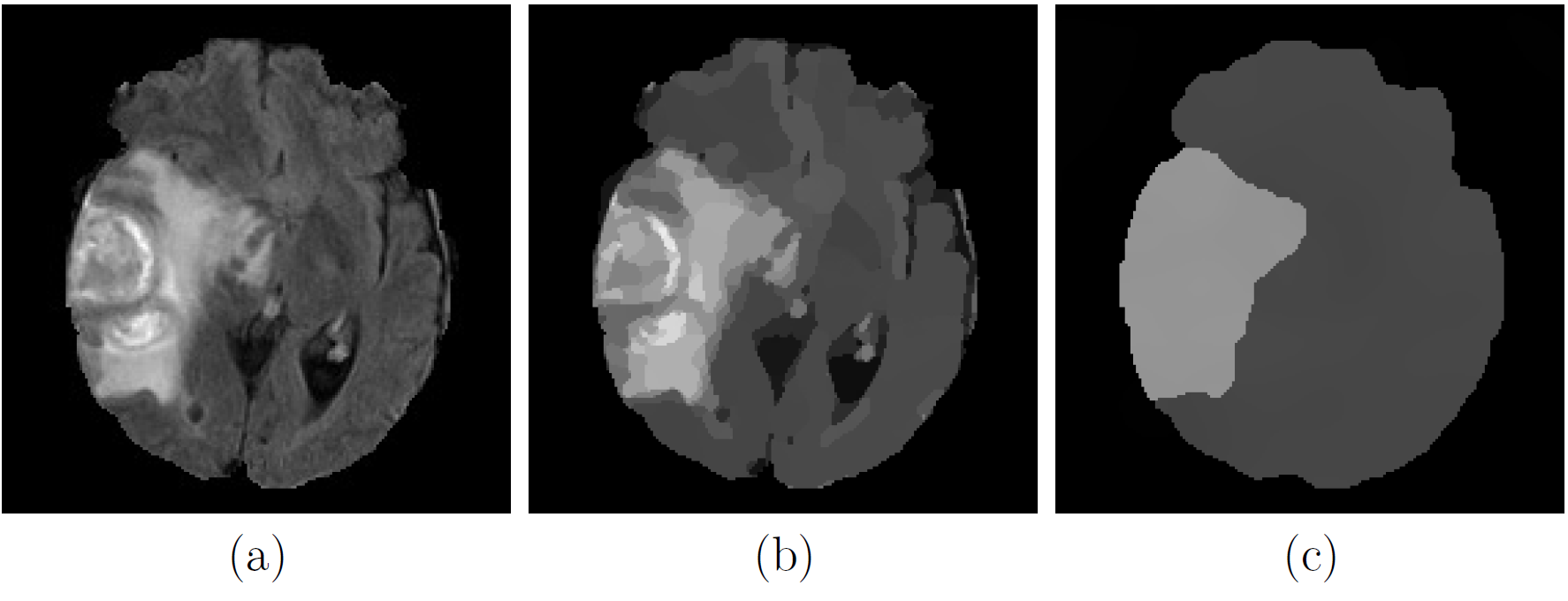}
  \end{center}
  \caption[Results of the Mumford-Shah function for different 
  penalizations.] 
  { 
Exemplary results for different penalizations of the boundary 
length. 
\textbf{(a)} $T_2$-Flair input image.
\textbf{(b)} Result for $\nu= 1000$.
\textbf{(c)} Result for $\nu= 340000$.
  }
  \label{fig:boundaryweight}
\end{figure}

Let us consider a cubic data domain $\Omega \subset \mathbb{R}^3$ and
some volumetric data set $\bm{f}: \Omega \rightarrow \mathbb{R}^m$. 
For our application, its $m$ channels describe different MRI modalities
such as $T_1$, $T_{1c}$, $T_2$ and $T_2$-Flair.
Then a segmentation of $\bm{f}$ by means of the Mumford--Shah cartoon 
model \cite{MuS85,mumford1989optimal} minimizes the energy functional
\begin{equation}\label{eq:mumford-shah}
E(\bm{u}, C) = \sum_i \int_{\Omega_i} \|\bm{u}\!-\!\bm{f}\|^2 d\bm{x} 
+ \nu \, \ell(C).
\end{equation}
Here the a priori unknown number of segments $\Omega_i$ partition the 
data domain $\Omega$, the function $\bm{u}$ denotes a piecewise constant 
approximation of $\bm{f}$, $\|\,.\,\|$ is the Euclidean norm in $\mathbb{R}^m$,
and the segment boundaries $C$ have a (Hausdorff) length of
$\ell(C)$. The first term of the energy is a data term that penalizes 
fluctuations within each segment, while the second term favors short 
segment boundaries. 
The parameter $\nu > 0$ allows to weight the boundary length in
relation to the inhomogeneities within each segment. Obviously the 
choice of $\nu$ is of crucial importance: The higher the value of 
this parameter, the less segments are contained in the final result. 
In Fig.~\ref{fig:boundaryweight}, the number of segments decreases with 
increasing penalization of the boundary length. At the same time, the 
inhomogeneities within individual segments increases.
On MRI $T_2$-Flair scans, high-grade gliomas contain areas that are 
brighter than the brain tissue. We use this prior knowledge and segment 
for a bright outlier in intensity in the following way: We start with the parameter $\nu=400,000$ and check if this gives a segmentation into two areas: 
the tumor and the background. If the area of the thresholded tumor is larger than 50\% of the brain volume, this is an indication that $\nu$ was too large such that the tumor has been merged with its background. In this case, we reduce $\nu$ by $15$\% and start the procedure again. This approach is repeated 
recursively until we have a segmentation where the tumor volume is below $50$\% 
of the brain volume.\\
We use this first segmentation to determine further tumor 
subcomponents: We minimize the Mumford-Shah cartoon model again with 
a very small boundary penalization ($\nu=1$), but this time exclusively 
on the $T_{1c}$ scans and in the previously defined segment. Afterwards 
we use Otsu's thresholding to identify the active tumor, i.e. 
enhancing- and non-enhancing tumor core. In this way, we get a 
splitting of the complete tumor region into active tumor 
and necrosis/edema. To get the final subcomponents, we apply Otsu's method on 
both subcomponents and split the first component, i.e.~active tumor, into its
enhancing and non-enhancing part and the second subcomponent into necrosis and
edema.


\subsubsection{U-Net}
\label{sec:U-Net}
\label{sec:dnn_U-Net}
\begin{figure}[b!]
  \hspace*{-5mm}
  \centering
  \begin{tabular}{c} 
\includegraphics[width=1\textwidth]{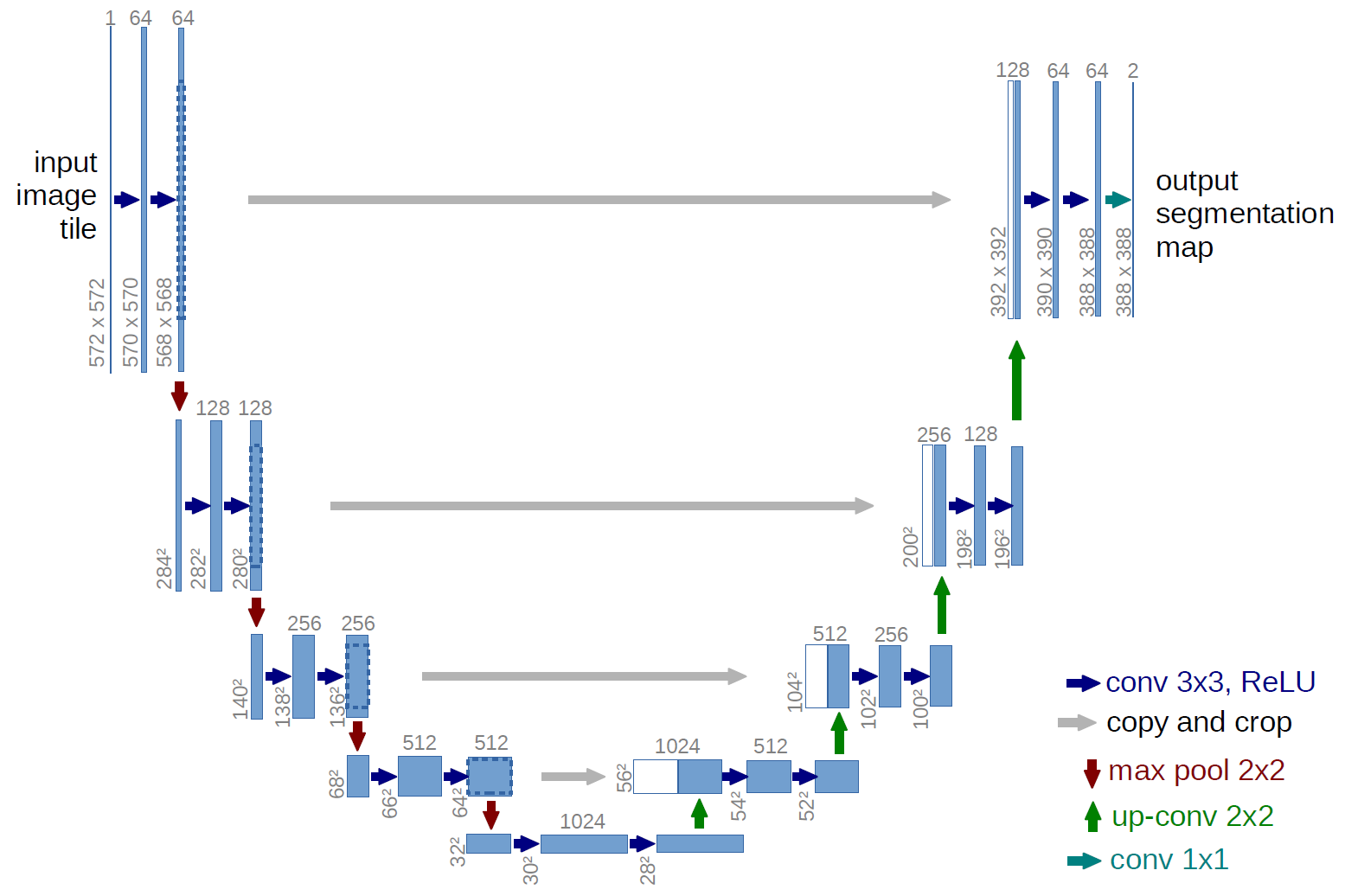}
  \end{tabular}
  \caption[Basic structure of U-Net approaches.]
  {Basic structure of U-Net approaches. Each blue box indicates a 
multidimensional feature map, arrows correspond to operations. Image 
courtesy of Ronneberger et al. \cite{RFB15a}.}
  \label{fig:U-Net_struc}
\end{figure}
The U-Net architecture \cite{RFB15a} is probably the most dominant topology in 
image segmentation. The intuition behind its structure is to re-use already learned feature 
mappings: This architecture can be split into two components, an 
encoder and a decoder branch connected by a bottleneck; see 
Fig.~\ref{fig:U-Net_struc}. While the first one learns feature mappings and 
contracts the image to its vector representation in the latent space (i.e. the 
bottleneck), the decoder part reconstructs an image of the original size using 
the previously learned feature maps, see \cite{RFB15a} for more 
details. In this way, the structural integrity is maintained while 
distortions due to lost locality are reduced. 
In order to introduce locality to the massively abstracted feature 
representations, Ronneberger et al. \cite{RFB15a} apply 
\textit{skip-connections}\index{Skip Connection}, allowing a re-usage of 
already learned filters.\\
In its original formulation, the U-Net architecture was developed for 2D 
cell images. However, its extension to 3D images (necessary for volumetric 
MRI data) is straightforward - the architecture is identical and only replaces 
all 2D operators with their corresponding 3d variants \cite{cciccek20163d}.\\
In the following we will discuss the No NewNet topology\cite{isensee2018nnu}. 
This recent work shows a very high performance on several datasets. It was 
developed on the basic assumption that already the original U-Net architecture 
is very powerful and most extensions of its design are not necessary and too 
complicated. Since we will follow this assumption in a quite similar way, we 
explain this work in detail.
\subsubsection{No NewNet}
\label{sec:nnU-Net}
Since the publication of the U-Net architecture, the encoder-decoder 
strategy has become the dominant approach in image segmentation. 
Nowadays, almost all new developments in this field are based on architectural 
modifications of this topology \cite{jegou2017one,isensee2018nnu,myronenko20183d}. \\
In the meantime it is almost impossible to predict which architecture 
might be suitable for a problem due to the multitude of possible extensions: 
Each of these possibilities has been tested on a specific data set. 
Unfortunately, it is an inherent part of deep learning that there is an 
architectural overfit to the data set used - making it almost impossible to 
decide whether an adjustment is appropriate in a different context.\\
Isensee et al. \cite{isensee2018nnu} implemented a number of these 
variants and evaluated their usefulness. It is not surprising that they found most 
of these extensions to be pointless in a general context - compared to a well 
trained U-Net model. Overall, they claim that a generic U-Net architecture with a 
few minor modifications can be sufficient to provide competitive 
performance.
\begin{figure}[b!]
  \hspace*{-5mm}
  \centering
  \begin{tabular}{c} 
\includegraphics[width=1\textwidth]{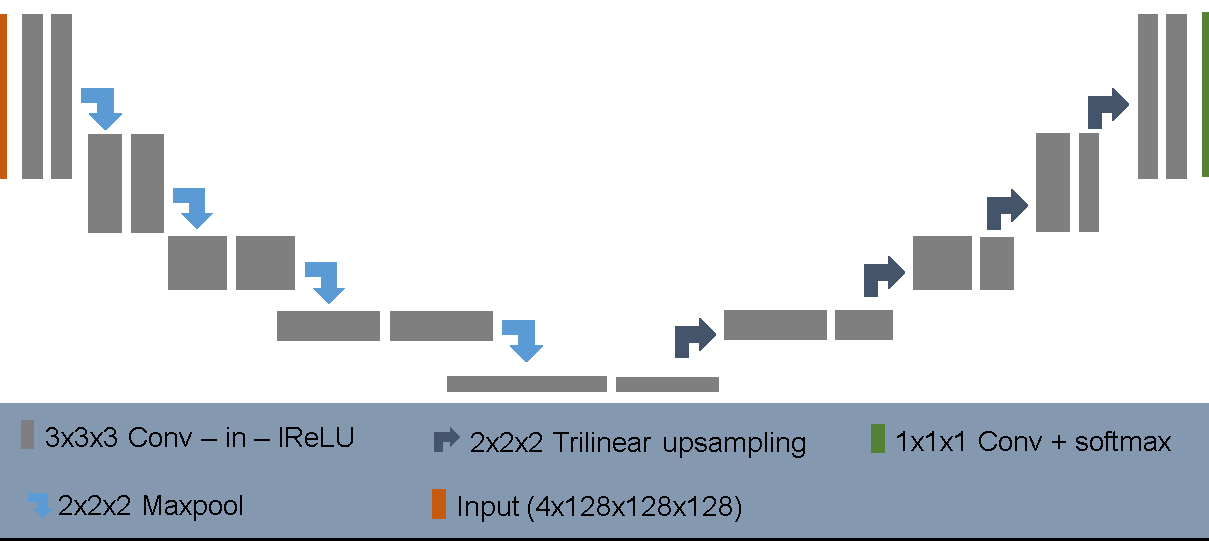}
  \end{tabular}
  \caption[Architecture of the No NewNet model.]
  {Architecture of the No NewNet model. Each gray box corresponds to 
  a series of convolution, instance normalization, and leaky relu. 
Arrows indicate up and down sampling operations, respectively. Image 
courtesy of 
Isensee et al. \cite{isensee2018nnu}.}
  \label{fig:nnU-Net_struc}
\end{figure}
\\Probably the only significant difference to the original scheme is a 
normalization after each convolution layer; see Fig.~\ref{fig:nnU-Net_struc}. 
Obviously, this is fully consistent with current findings that normalization 
lead to wider optima (with higher generalization 
performance) in the loss surface \cite{izmailov2018averaging}.
In order to optimize the performance of the model on BraTS benchmark 
data, the authors suggest a set of additional extensions, see 
\cite{isensee2018nnu} for more details.
All in all, each of those steps contributed some improvement to the 
overall performance \cite{isensee2018nnu}. While most of their adjustments had 
only minor influence on the segmentation performance, the 
postprocessing step as well as the training on additional data noticeably 
improved the error metric by $0.032$ (enhancing core) and $0.013$ (complete 
tumor), respectively. This indicates that their main improvement 
in performance was caused by the inclusion of more training data, i.e. by 
reducing the overfit of the model to the training distribution.
\subsubsection{NVDLMED: Autoencoder Regularization}
\label{sec:nvdlmed}
\begin{figure}[t!]
  \hspace*{-5mm}
  \centering
  \begin{tabular}{c} 
\includegraphics[width=1\textwidth]{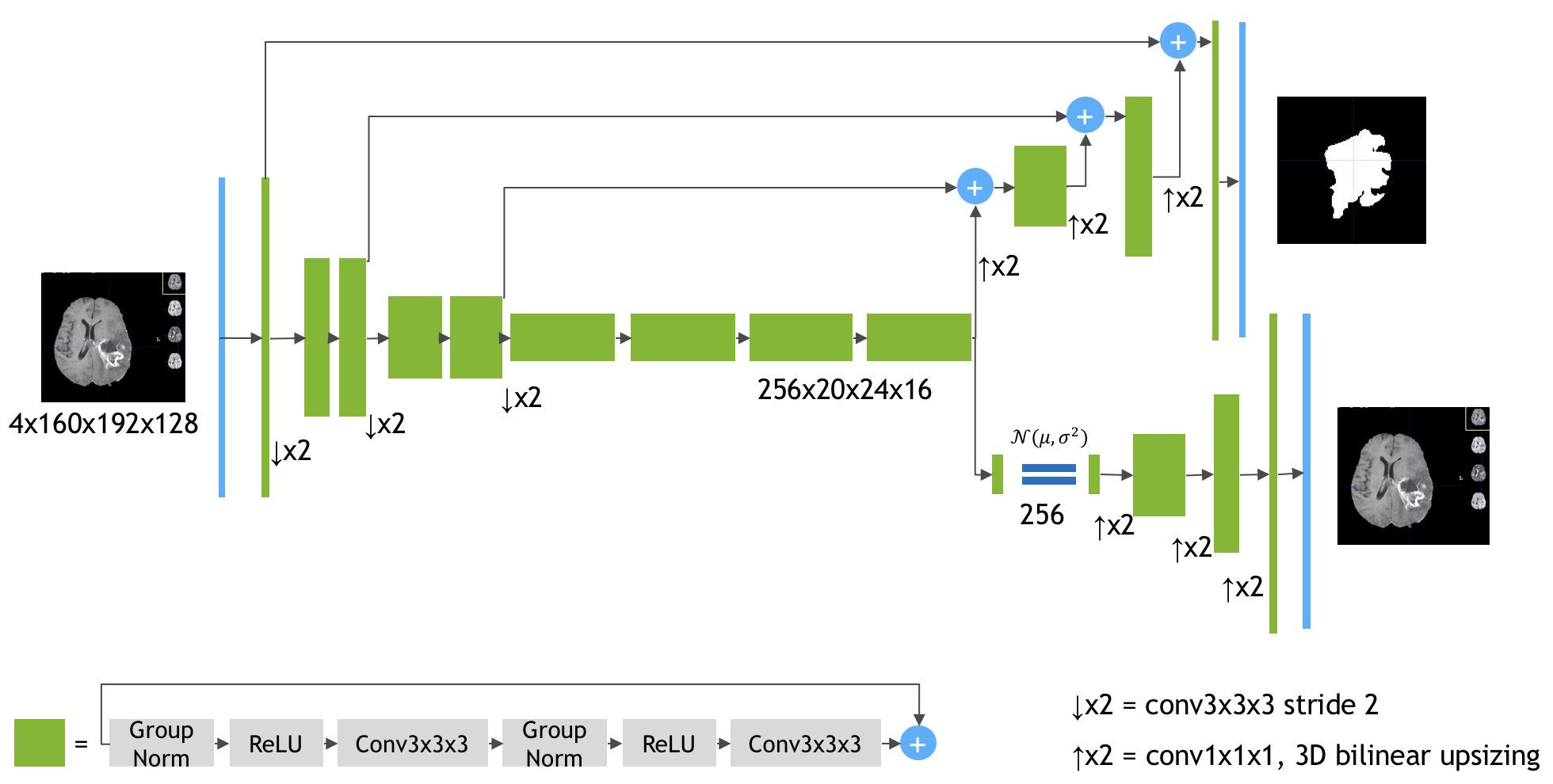}
  \end{tabular}
  \caption[Architecture of NVDLMED.]
  {Architecture of NVDLMED. In contrast to the basic U-Net structure, a second
  decoder branch is implemented.}
  \label{fig:3dmri_struc}
\end{figure}

The winner of the 2018 BraTS challenge also followed a basic U-Net architecture 
\cite{myronenko20183d}. While the backbone can still be reduced to an encoder-decoder 
structure, the author dramatically increased the model size and extended most 
of the basic topology by additional operations: Although the encoder branch 
is still similar, its building blocks are massively changed;  see 
Fig.~\ref{fig:3dmri_struc}. Probably the most important change is an additional 
variational autoencoder branch reconstructing the input image to itself. This 
sub-network is then used during the training phase as regularization.\\
In order to improve the model performance, NVDLMED is built on an ensemble of 
10 different networks. Unfortunately, this setting results in a very large 
network, that can only be trained on at least NVidia V100 GPUs, or on a CPU 
cluster.

\subsubsection{Cascadic Neural Networks}
\label{sec:cascnn}
\begin{figure}[t!]
  \hspace*{-5mm}
  \centering
  \begin{tabular}{c} 
\includegraphics[width=0.8\textwidth]{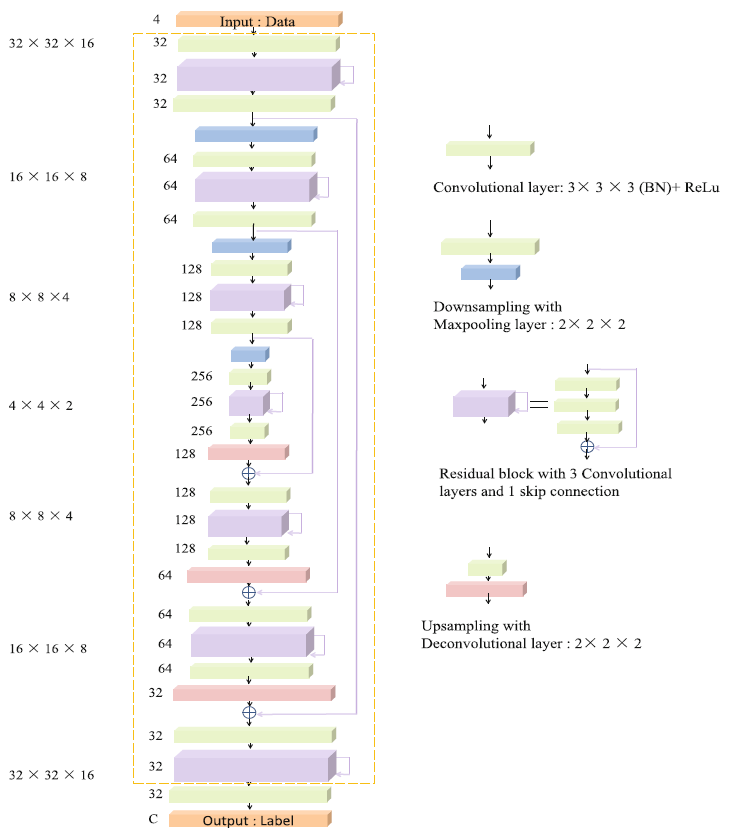}
  \end{tabular}
  \caption[Topology of FusionNets.]
  {Topology of FusionNets. Although the basic U-Net structure is still
  present, the architecture changed dramatically.}
  \label{fig:fusnet}
\end{figure}

Zhou et al. \cite{zhou2018learning} approach the task of brain tumor 
segmentation from a slightly different perspective. While most of the 
state-of-the-art methods consider the identification of the complete tumor and 
its subcomponents as a single problem, the authors decompose the segmentation 
challenge into three different sub-tasks. In a first step their method performs
a coarse segmentation to detect the complete tumor. Afterwards, the 
segmentation is refined and intra-tumoral classes are segmented. Finally, this 
segmentation is again optimized to classify the enhancing tumor core.
This cascade of segmentation tasks is realised with two different network
topologies. On the one hand, Zhou et al. make use of 3D FusionNets 
\cite{vidyaratne2018deep}; see Fig.~\ref{fig:fusnet}
to extract the multi-scale context information. On the other, they apply one-pass multi-task networks \cite{zhou2018one}.
In addition, Zhou et al. \cite{zhou2018learning} perform several modifications,
such that the final ensemble contains seven different neural network architectures whose results are averaged for the final model prediction.

\subsubsection{Preprocessing}
Typically MR images are recorded from different hospitals with varying 
scanners and no standardized parameter settings. This results in strong 
variations in the MR intensities: Even the same sequence of the same patient 
(e.g. $T_2$) acquired at the same machine, can differ dramatically due to 
inconsistent parameter choices. Deep neural networks learn the data 
distribution provided by the training set. Hence, it is essential that the 
value range in the training data corresponds to the range present in the test 
set. In order to compensate for these variations, we follow 
\cite{isensee2018nnu} and adjust each modality independently. In a first step, 
we substract the mean of the brain region and normalize by its standard 
deviation. Afterwards, we remove outliers by clipping and rescale the images to 
the range $[0,1]$.

\subsubsection{Postprocessing}
\label{sec:ec:post}
Although deep neural networks proved to produce segmentation results of 
high quality, post-processing is a necessary step in a medical context. The 
brain tumor segmentation challenge contains high- and low-grade gliomas. 
While the high-grade tumors typically consist of an enhancing tumor core, it is 
rarely present in low-grade abnormalities. \\
In order to compensate for this prior knowledge, we follow \cite{isensee2018nnu}
and apply a postprocessing step to remove potentially false labels of the 
enhancing tumor core in low-grade gliomas. \\
Our interactive segmentation approach \cite{mueller2016} is a recent 
method with good results on kidney tumors \cite{muller2019benchmarking} 
and already proved its ability to correct for false labels. 
Since it is less well-known, let us discuss it in more detail.
We follow \cite{chambolle2012convex} and consider a minimal 
partitioning problem of the cubic image domain $\Omega \subset \mathbb{R}^3$ 
into  $\Omega_1,\ldots,\Omega_n\subset\R^3$ non-overlapping regions
\begin{equation} \label{eq:min-part}
\begin{split}
\min_{\Omega_1,\ldots,\Omega_n\subset\Omega}\, \frac 12 \sum_{i=1}^n 
\Per(\Omega_i; \Omega) + \sum_{i=1}^n \Int{\Omega_i} {h_i(\x)} \x \,,\\
\text{s.t. } \quad \Omega = \bigcup_{i=1}^n \Omega_i, \quad \Omega_i \cap 
\Omega_j = \emptyset, \quad \forall~i\neq j
\end{split}
\end{equation}

where $\Per(\Omega_i; \Omega)$ denotes the perimeter of region $\Omega_i$ 
inside $\Omega$, and $\map{h_i}{\R}{\R_+}$ are potential functions reflecting 
the cost for each pixel being assigned to a certain label $i=1,\ldots,n$. To 
align image and region boundaries, the perimeter is commonly measured in a 
metric induced by the underlying image $\bm{f}: \Omega \rightarrow 
\mathbb{R}^3$. In this application, we weight the perimeter $\Per_g(\Omega_i; 
\Omega)$ of region boundaries in the metric

\begin{equation*}\label{eq:perimeter_unger}
{g(\x) = \exp({-\edge(\x)^{\beta}}/\,{\bar{\edge}}) \,, \quad 
  \bar{\edge}:=\frac{2}{|\Omega|}\Int {\Omega}{\abs{\edge(\x)}}\x 
}.
\end{equation*}
Here $\map{\edge}{\Omega}{\R}$ is the output of the fast structured 
edge detector of \cite{dollar2013structured,dollar2015fast} and 
$\beta$ is a positive parameter. Assume a (measurable) set of 
user-scribbles 
$\scrib_i\subset 
\Omega$ for each 
label $i$ is given. We define the potential functions $h_i(\x)$ in 
(\ref{eq:min-part}) as the negative logarithm of
\begin{equation}\label{eq:cost_func}
\begin{split}
&{\tilde h_i(\x)} = 
\begin{cases}
\left\lbrace\frac{1}{\abs{\scrib_i}} \Int {\scrib_i} 
{G_{\rho}~G_{\sigma}} \y\right\rbrace_{\text{scale}},  
&\x\notin\scrib_j,\\
1-\zeta,  &\x\in\scrib_j, i=j, \\
\zeta/(n-1),  &\x\in\scrib_j, i \neq j,
\end{cases}
\end{split}
\end{equation}
and
\begin{equation*}
\begin{split}
& G_{\rho} = k_{\rho_i(\x)}(\x - \y),\\
& G_{\sigma} = k_{\sigma}(\img(\x)- \img(\y)).
\end{split}
\end{equation*}
Here $\{.\}_{\text{scale}}$ denotes linear rescaling to 
$[0,1]$, $\abs{\scrib_i}$ is the area occupied by $i$th label, $\zeta$ 
is the assumed probability for a scribble being correct, and $k_\sigma$ and 
$k_{\rho_i}$ are Gaussians with standard deviation $\sigma$ in intensity space 
and adaptive standard deviation $\rho_i(\x) = \alpha \inf_{\y\in\scrib_i} 
\abs{\x - \y}$ in the spatial domain, respectively. The spatially adaptive 
standard deviation attenuates the influence of the intensity distribution from 
scribbles that are far away proportionally to the distance of $\x$ to the 
closest scribble location. 
Hence, we postprocess the segmentation masks of the deep learning 
models as follows: In a first step, we sample every eight voxel in the output 
mask to sparsify the data. Afterwards we incorporate this mask in the cost term
of our semi-automatic approach and densify the segmentation.

\subsection{Improving the Generalization Performance}
Overfitting is one of the major problems in training of deep neural 
networks. Typically, this issue is caused by a lack of training data in 
combination with complex models. Especially in the situation of medical image 
segmentation, the amount of data is rather limited. There are several 
approaches to relax this problem: Obviously, the most straight forward idea is 
to add more training data. However, this is typically a severe problem. 
Data augmentation is therefore often used to circumvent the 
lack of further 
data. In this approach, additional data is simulated by random rotations, 
intensity shifts, axis mirror flips, or the addition of noise distributions, 
for example. While almost all current methods use data augmentation, the 
simulation of different noise distributions is generally not used. In our 
opinion, this has two reasons: First, it is not possible to include every 
distortion that occurs.  Although DNNs can handle the exact distortion they 
were trained on perfectly, they nevertheless show a strong generalization 
failure towards previously unseen variations \cite{geirhos2018generalisation}.
However, the overfit to a specific dataset is reduced, so 
that although a better generalization can be achieved, the overall performance 
on the dataset drops slightly.
Another possibility is to reduce the capacity of a model by reducing its size. 
Of course, it is also an option to regularize either the weights or the loss 
functions of a model. One more strategy is to include normalization layers: 
Recent work \cite{izmailov2018averaging} indicates that normalization 
layers lead to wider optima and therefore better generalization.\\
In the following, we discuss two approaches: The first one, octave 
convolutions \cite{chen2019drop}, addresses the reduction of weights in a 
neural network while not reducing its capacity. This advanced operator allows 
to exploit the mixture of frequencies inherent to each image. Second, we 
illustrate the stochastic weight averaging \cite{izmailov2018averaging} 
that enables the optimization algorithm to converge to wider and therefore 
better generalizing optima in the loss surface.
\subsubsection{Octave Convolutions}\label{sec:octconv}

\begin{figure}[b!]
  \hspace*{-5mm}
  \centering
\includegraphics[width=0.9\textwidth]{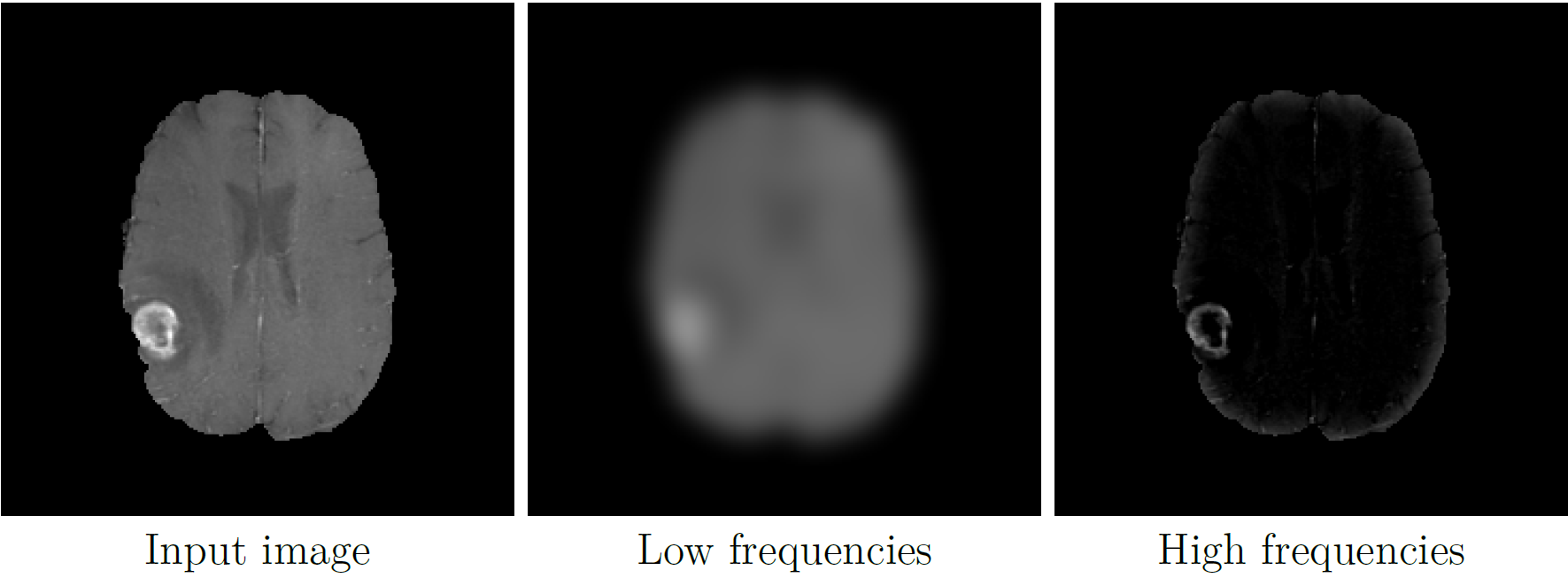}
  \caption[Illustration of low and high frequency parts in an image.]
  {Illustration of low and high frequency parts in an image. The 
  input image of an MRI of the brain (left) is split into its low frequencies 
(middle) and high frequencies (right).}
  \label{fig:octconv_idea}
\end{figure}

\begin{figure}[b!]
  \hspace*{-5mm}
  \centering
  \begin{tabular}{c} 
\includegraphics[width=0.4\textwidth]{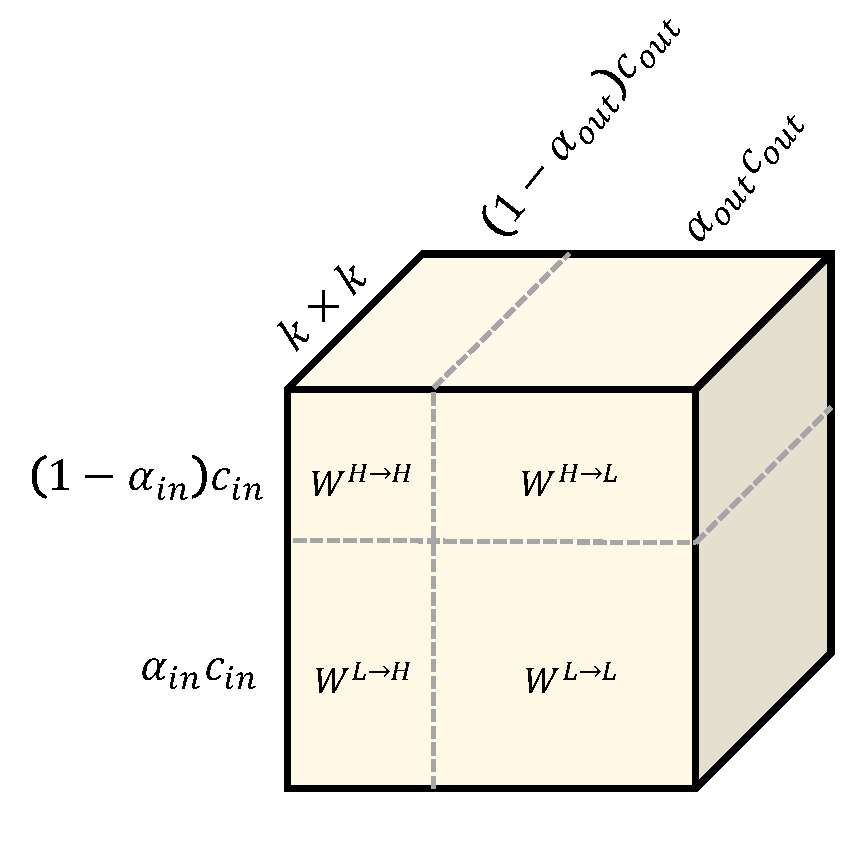}
  \end{tabular}
  \caption[Illustration of the octave convolution kernel.]
  {Illustration of the octave convolution kernel. The kernel is split 
  in 
intra- and inter-frequency parts. Image courtesy of Chen et 
al. \cite{chen2019drop}.}
  \label{fig:octconv_kernel}
\end{figure}

The fundamental aspect of convolution layers is their ability to 
identify local structures in their input data. These characteristics are then assigned 
to a new filter response - typically the image resolution does not change 
during this process. \\
However, each image can be divided into its low-frequency signal, which 
describes the coarse structure and the global layout, and its high-frequency 
signal, containing fine details; see Fig.~\ref{fig:octconv_idea}. Although this 
is well known in the classic image processing community, this inherent 
information cannot be exploited by standard convolutional layers. Recently 
there are several attempts to express this structure within layers of deep 
neural networks \cite{ke2017multigrid,chen2019drop}. The multigrid approach of 
Ke et al. \cite{ke2017multigrid} maps every convolutional layer into a pyramid 
of operations. In this way, features at different scales can be extracted. 
However, this type of strategy obviously has a massive disadvantage: The amount 
of required parameters increases with the number of scales in the pyramids. \\
\begin{figure}[t!]
	\hspace*{-5mm}
	\centering
	\begin{tabular}{c} 
		\includegraphics[width=1\textwidth]{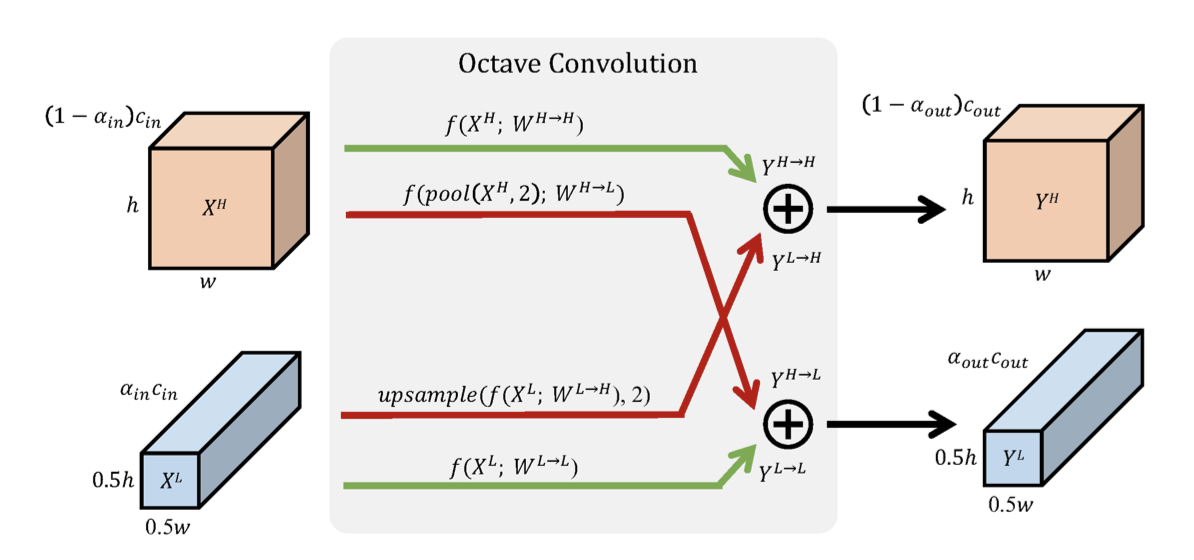}
	\end{tabular}
	\caption[Detailed design of octave convolutions.]
	{Detailed design of octave convolutions. Red arrows indicate 
		communication 
		between low- and high-frequency components, green arrows depict 
		regular 
		information updates. Image courtesy of Chen et al. 
		\cite{chen2019drop}.}
	\label{fig:octconv_scheme}
\end{figure}
\textit{Octave convolutions}\index{Octave Convolution} use a similar concept 
but interpret output feature maps as mixtures of information at different 
frequency scales \cite{chen2019drop}. Hence, these advanced convolutions 
factorize the output maps only into two groups: low and high 
frequencies. The corresponding smoothly changing low-frequency maps are 
then stored in a low resolution tensor (half of the original input resolution) 
to reduce spatial redundancy \cite{chen2019drop}; see 
Fig.~\ref{fig:octconv_scheme}. \\
Following this idea, octave convolutions process low frequency information with 
corresponding (low frequency) convolutions. This not only increases the 
receptive field in the original pixel space, but also collects more contextual 
information. Since the resolution for the low-frequency filter responses can be 
reduced, this saves both computational load and memory consumption. \\
The effort for such an octave convolution architecture consists of an 
additional hyperparameter $\alpha \in [0; 1]$ indicating the ratio of 
low frequency components. In order to compute the output feature maps, the 
convolution kernel is split accordingly; see Fig.~\ref{fig:octconv_kernel}.
\\Obviously, filter responses of intra-frequency maps can be computed 
with regular convolutions. However, up- and down-sampling (or pooling) 
operations for inter-frequency computations can also be folded up into the 
convolutions; see \cite{chen2019drop} for more details.
\\In total, the application of octave convolutions is straight-forward. 
Due to its inherent design, it is a plug-and-play component, not leading to 
any architectural changes. In some of our experiments, we replaced 
all standard convolutions with their octave variants. Although this change 
had no consequences with respect to network architecture, it dramatically 
reduced the memory footprint of the models as well as training time per epoch 
while improving their generalization behavior; see Sec.~\ref{dnn:experiments}.

\subsubsection{Stochastic Weight Averaging}
\label{sec:swa}
\begin{figure}[b!]
  \hspace*{-5mm}
  \centering
  \begin{tabular}{c} 
\includegraphics[width=1\textwidth]{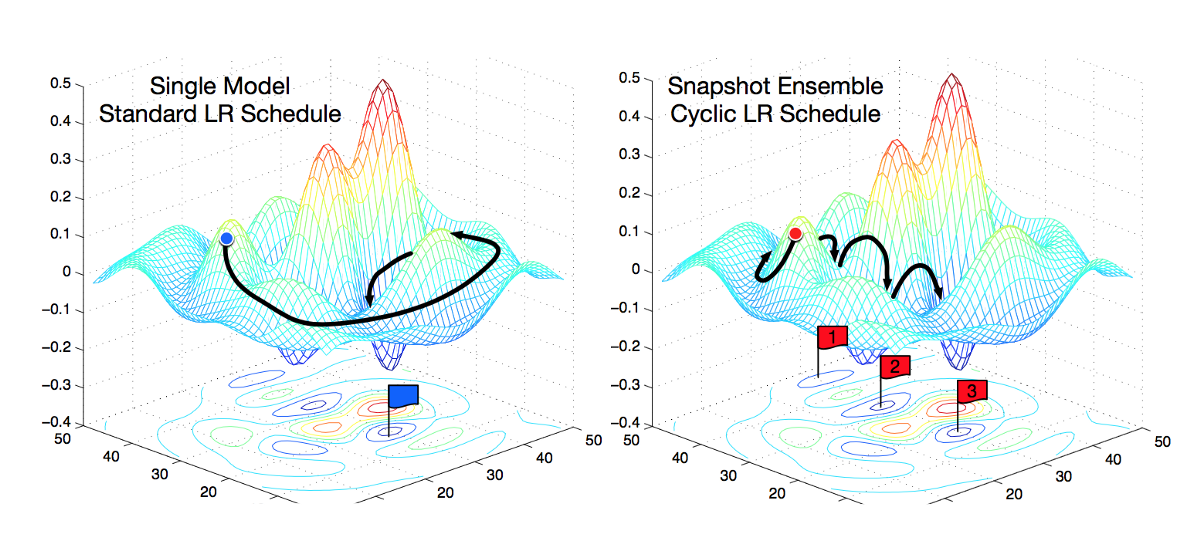}
  \end{tabular}
  \caption[Illustrations of different Model Snapshots.]
  {Illustration of different model snapshots. While the standard 
  learning rate schedule slowly converges to the minimum, snapshot 
ensembles are a combination of different local optima. Image courtesy of Huang 
et al. \cite{HuangLP0HW17}}
  \label{fig:snapshot}
\end{figure}

The training of deep neural networks is a tedious and time consuming 
task. While in most cases, the capacity of the model architecture is 
large enough to solve the depicted problem, finding reasonable 
hyperparameters (e.g. learning rate, batch size, etc) can be challenging: 
Especially the learning rate has massive influence to the training procedure 
and an optimal value is of crucial importance. In medical image segmentation, 
neural network architectures tend to be complicated and can easily overfit due 
to a limited amount of training data. In this scenario, an appropriate learning 
rate is even more important. \\
Typically, deep neural networks do not converge to a global minimum. 
Therefore, the quality of the model is evaluated with respect to its 
generalization performance. In general, local optima with flat basins tend to 
generalize better than those in sharp areas 
\cite{smith2017cyclical,HuangLP0HW17,izmailov2018averaging}. Since even 
small changes of the weights can lead to dramatic changes in the model 
prediction, these solutions are not stable. If the learning rate is too low, 
the model converges to the nearest local optimum and may hang in a sharp basin. 
Once the learning rate is high enough, the inherent random motion of the 
gradient steps not only prevents the solution from being trapped in one of the 
sharp regions, but can also help the optimizer to escape. Obviously, finding a 
reasonable learning rate boils down to the trade-off between convergence and 
generalization.\\
Probably the most common strategy to solve this problem is the usage of 
an cyclic scheme \cite{smith2017cyclical,loshchilov2017sgdr}.
In \textit{cosine annealing}, the learning rate cyclically decreases 
from a given maximal value following the cosine function 
\cite{loshchilov2017sgdr}. It turned out, that each of the local optima at the 
end of the cycles had similar performance, but lead to different but not 
overlapping errors in the model prediction; see Fig.~\ref{fig:snapshot}. 
Hence, Huang et al. \cite{HuangLP0HW17} suggested to combine the 
local optima of each cycle into an ensemble prediction. Unfortunately, 
computation time at inference increases dramatically with the number of 
snapshot models used in the ensemble. \\
\textit{Stochastic weight averaging} follows the same idea but at a 
fraction of computational load. The basic idea is to conduct an equal average 
of the weights traversed by the optimizer with a learning rate schedule 
\cite{izmailov2018averaging}.
Intuitively, by taking the average of several local optima in the loss 
surface, a wider basin can be reached with better generalization performance 
\cite{izmailov2018averaging,athiwaratkun2018there}.\\
In contrast to ensemble approaches, we only need two models: The first 
one keeps track of the running average of the model weights, while the 
second one is traversing the weight space.
At the end of each learning rate cycle, the state of the second model 
is used to update the weights of the running average model as

\begin{align}
w_{\text{swa}} = \frac{w_{\text{swa}}\cdot n_{\text{models}}+ 
w}{n_{\text{models}}+1}.
\end{align}
\noindent
Here, $w_{\text{swa}}$ are the weights of the running average model, 
while $w$ are the weights of the model traversing the weight space, 
respectively. The total number of models to be averaged is given by 
$n_{\text{models}}$.
\begin{figure}[t!]
  \hspace*{-5mm}
  \centering
\includegraphics[width=0.95\textwidth]{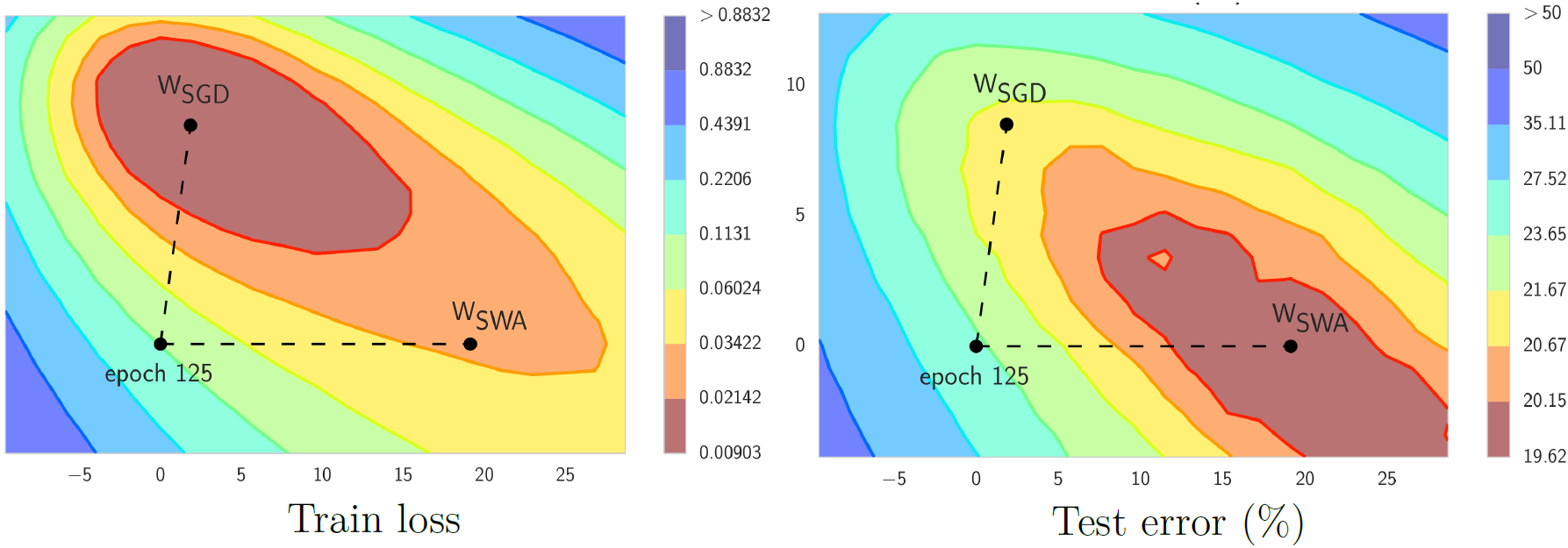}
  \caption[Illustrations of SWA and SGD.]
  {Illustration of SWA and SGD showing the weights suggested by SGD 
  and SWA at convergence. SWA started the weights of SGD after 125 training 
  epochs. Image courtesy of Izmailov et al. \cite{izmailov2018averaging}}
  \label{fig:losssurf_swa}
\end{figure}
All in all, stochastic weight averaging significally improves generalization 
performance \cite{athiwaratkun2018there}, being less prone to the shifts 
between train and test error loss; see Fig.~\ref{fig:losssurf_swa}.
\begin{figure}[b!]
  \centering
\includegraphics[width=0.65\textwidth]{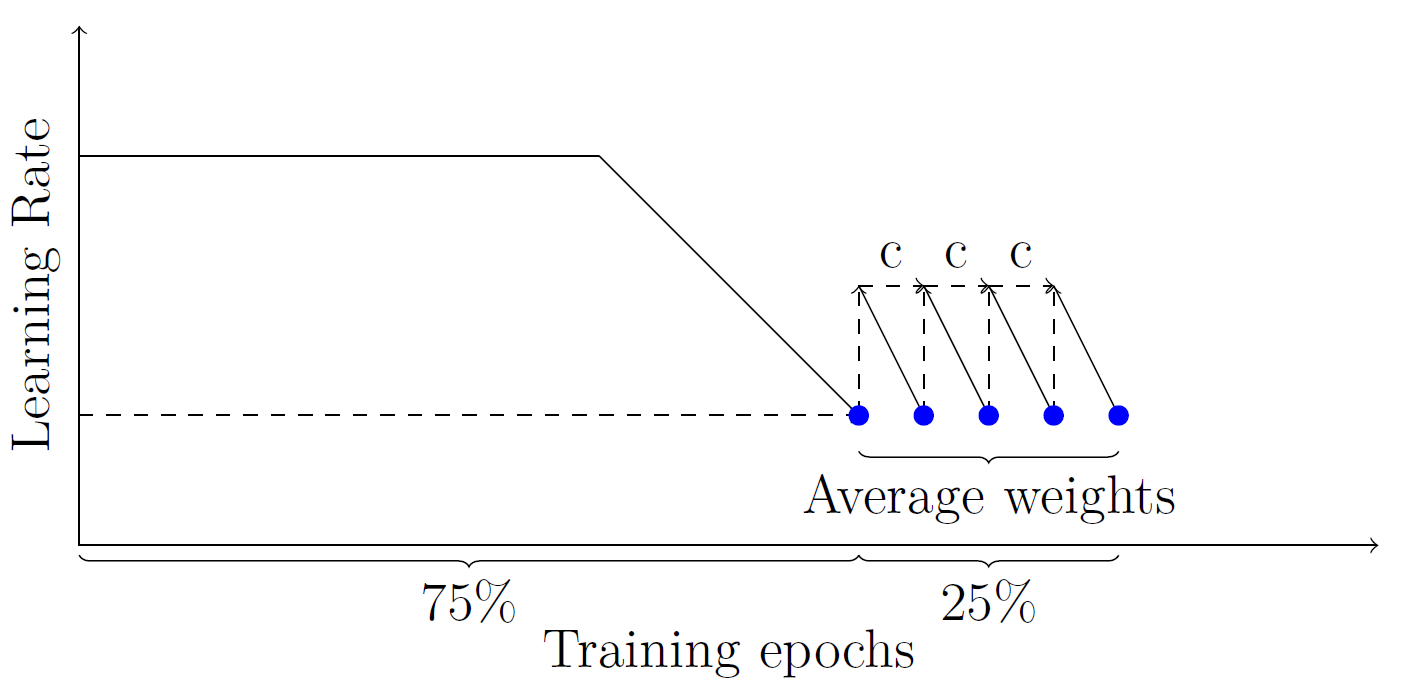}
  \caption[Illustration of stochastic weight averaging.]{Sketch of 
  stochastic weight averaging.}
  \label{fig:swa_sketch}
\end{figure}
In general the strategy can be divided in two phases: In the first 
phase of $75\%$ of training time, the learning rate schedule follows a standard 
scheme - e.g. it is fixed to a specific value and decays after several epochs. 
In the second phase, the learning rate can be set to a constant value or 
follow a cyclic scheme to encourage the exploration of the loss surface; see 
Fig.~\ref{fig:swa_sketch}.

\section{Results}
\label{dnn:experiments}
\begin{figure}[t!]
\hspace*{-5mm}
\centering
\includegraphics[width=0.95\textwidth]{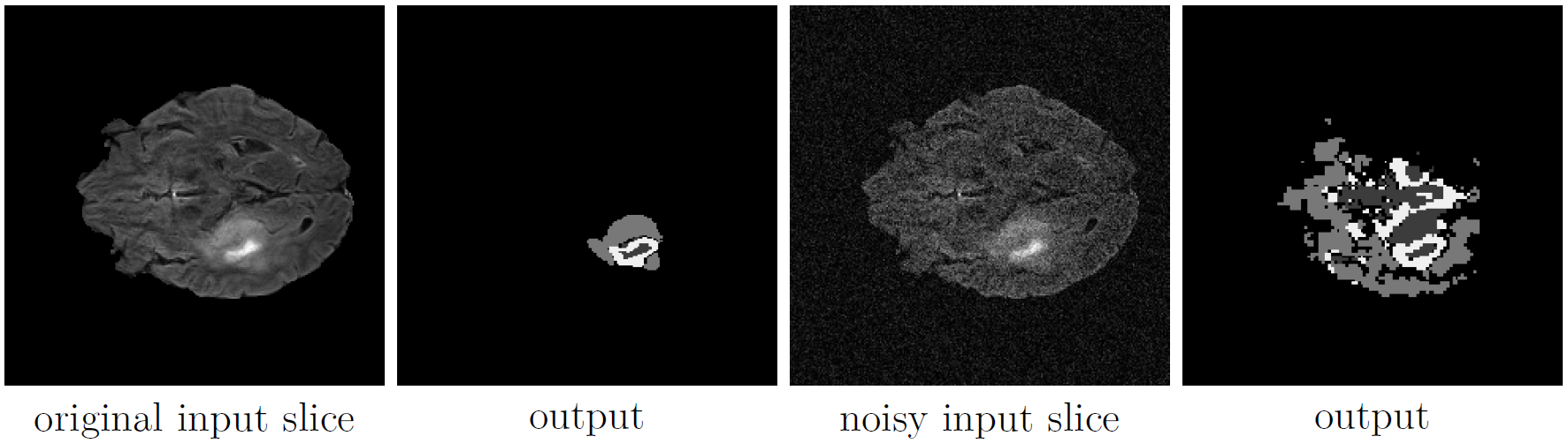}
\vspace{2mm}
	\caption[Exemplary performance drop of DNNs on noisy data.]
	{Exemplary result for the NoNewNet model on original, and slightly 
	disturbed ($\sigma=0.02$) input data. }
	\label{fig:ex_noise_seg}
\end{figure}

The brain tumor segmentation challenge is a widely accepted benchmark data set 
\cite{menze2014multimodal,bakas2017advancing,bakas2018identifying}. 
The challenge contains skull-stripped and spatially registered multimodal MR 
images ($T_1$, $T_{1c}$, $T_2$, and $T_2$-Flair) with a voxel size of 
$1$\textit{mm} in every direction. Tumors are of different shape, size and 
location in each data set.\\
In 2018, the BraTS challenge contained 285 training instances accompanied with 
66 validation and 191 test cases. Unfortunately, the testing data set allows 
only for a single submission, disqualifying this compound for our analysis. 
However, we found the validation data set to be rather small and therefore not 
expressive. We decided to rely in our evaluation on five-fold cross validation 
on the training data set. We follow Isensee's assumption \cite{isensee2018nnu}, 
that the conclusions drawn from the training set with cross 
validation are more general in nature and more robust to changes in the 
underlying distributions.\\
We performed nearly all network training on four NVidia Titan V with 12GB 
memory and 5120 cuda cores. In case of NVDLMED, we do not have a graphics card 
with sufficiently large memory: We trained this network for several weeks on  
Intel Xeon Gold 6132 (\enquote{Skylake}) with 28 CPU cores and 192GB of main 
memory. \\
We set all hyperparameters of the considered networks as described in their 
publications and used code provided by the authors whenever possible.\\
In order to generate a baseline for our experiments, we evaluated all analyzed
approaches on the original BraTS2018 training data; see 
Tab.~\ref{tab:brats_orig}. Here, \enquote{CMS} denotes our cascadic 
Mumford-Shah method (Sec.~\ref{sec:cms}) while \enquote{CascNN} means the 
cascadic segmentation approach with multiple neural networks 
(Sec.~\ref{sec:cascnn}), and \enquote{No NewNet} refers to the No NewNet 
approach with region optimization and postprocessing (Sec.~\ref{sec:nnU-Net}). 
Please note that \enquote{NVDLMED} represents a single NVDLMED network 
(Sec.~\ref{sec:nvdlmed}) not an ensemble of several models.
\begin{table}[t!]
  \caption[BraTS18 evaluation for different segmentation approaches.]{BraTS18 
  evaluation for different segmentation approaches in terms of Dice score. No 
  additional disturbances.}
  \label{tab:brats_orig}
  \begin{center} 
    \begin{tabular}{l|c|c|c}
      \hline
      \noalign{\smallskip}
      Method & Enhancing & Complete  & Core\\
      \noalign{\smallskip}
      \hline
      \noalign{\smallskip}
      CMS   & $0.70$ & $0.84$ & $0.76$\\
      U-Net~ & $0.73$ & $0.89$ & $0.82$\\
      CascNN~ & $0.78$ & $0.89$ & $0.84$\\
      No NewNet~ & $0.77$ & $0.90$ & $0.84$\\
      NVDLMED~ & $0.82$ & $0.91$ & $0.86$\\
      \noalign{\smallskip}
      \hline
    \end{tabular}
  \end{center}
\end{table}
\\The neural networks surpass the cascaded Mumford Shah approach as expected. 
Nevertheless, the assumption that a brain tumor has higher average intensities 
in $T_2$-Flair images is a reliable prior knowledge: This intuitive 
method shows a remarkable performance when the entire tumor is considered. 
However, the most significant difference between the results of the 
various networks is shown in their accuracy to identify the enhancing tumor 
core.\\
\begin{table}[b!]
  \caption{BraTS18 evaluation for different segmentation approaches. Gaussian 
    Noise ($\sigma=0.02$) is added to the validation data.}
  \label{tab:brats_noise_val}
  \begin{center} 
    \begin{tabular}{l|c|c|c}
      \hline
      \noalign{\smallskip}
      Method & Enhancing & Complete  & Core\\
      \noalign{\smallskip}
      \hline
      \noalign{\smallskip}
      CMS   & $0.69$ & $0.82$ & $0.74$\\
      U-Net~ & $0.71$ & $0.82$ & $0.75$\\
      CascNN~ & $0.65$ & $0.76$ & $0.76$\\
      No NewNet~ & $0.72$ & $0.83$ & $0.79$\\
      NVDLMED~ & $0.68$ & $0.80$ & $0.74$\\
      \noalign{\smallskip}
      \hline
    \end{tabular}
  \end{center}
\end{table}
\\Typically, a medical benchmark data set is intended as a biased version of a 
parti\-cular problem, i.e. in the case under consideration, all patients with 
high-grade brain tumors in MRI sequences. BraTS addresses this issue by 
providing comprehensive multi-institutional routine examinations of 
glioblastoma multiforme (GBM/HGG) and low grade gliomas (LGG) with 
pathologically confirmed diagnosis 
\cite{bakas2018identifying,bakas2017advancing}. However, care was 
mostly taken to create a representative visual representation of the brain 
tumors themselves. In a real clinical scenario, time and cost pressures usually 
prevail. For this reason, the assumption that voxels have a size of 
$1$\textit{mm} in all directions is not realistic. In fact, exactly the 
opposite is typically true: While in-slice images are taken at high resolution, 
across-slice images are mostly sampled at lower resolution. \\
In addition, noise also plays an important role in MRI images. These recordings 
are very costly and time-consuming: Often, MR sequences differ dramatically in 
sampling rates and suffer from heavy noise disturbances. All in all, real 
clinical MR images do not correspond to the scheme of the BraTS benchmark data.
In order to ensure the applicability of segmentation approaches tested on BraTS 
data in everyday clinical practice, it is necessary for them to show high 
generalization performance.\\
For this reason, we analyze the outcomes of the different approaches when the 
distribution of the validation data set does not exactly match that of the 
training data. In a first step, we add Gaussian noise with zero mean and 
standard deviation $\sigma=0.02$ to the validation data. The results are 
depicted in Tab.~\ref{tab:brats_noise_val}.
The Dice scores indicate that the prior information about tumor appearance used 
in the cascadic Mumford-Shah approach is highly robust to disturbances. 
Although this approach performed worse than the considered neural networks in 
the original setting, it copes relatively well with the noisy data and the Dice 
score is only marginally reduced ($\approx0.02$ for all categories).\\
On the contrary, all of the tested neural networks have a major problem with 
the different distribution in the validation data. All of them show a 
significant decline in their segmentation performance. This problem obviously 
also becomes more serious the more complicated the respective architecture is. 
While the basic U-Net as well as the No NewNet model drop by a Dice score 
of $\approx0.06$ on average, the much more complex CascNN and NVDLMED show a 
significant decline by a Dice score of $\approx0.11$ and $\approx0.12$, 
respectively; see Fig. \ref{fig:ex_noise_seg}.
\noindent
This is consistent with our assumption that the best performing models are not 
the ones that generalize best on the test data, but only have the strongest 
overfit. This conclusion unfortunately disqualifies models trained and 
evaluated on BraTS data to be directly applied in a real clinical scenario.\\
The two approaches No NewNet and NVDLMED were almost equal in the evaluation of 
the BraTS18 challenge and our analysis in Tab.~\ref{tab:brats_orig}. Since the 
NVDLMED in particular shows a strong overfit on the data set while its training 
is extremely computationally intensive, we exclude this network in the 
following from our evaluation. 
\\An obvious remedy to cope with noisy validation data is to add the same noise 
distribution to the training data as well; see 
Tab.~\ref{tab:brats_noise_valtrain}. In fact, this additional information helps 
the three deep learning approaches to handle the altered distribution and their 
performance returns close to the original value. Of course it would be possible 
to add different noise distributions to the training data. However, at training 
time it is usually not known how much noise is present in the test set. Another 
approach would be to include a preprocessing step to denoise the input images. 
Unfortunately, this idea also has a massive disadvantage: Small details might 
be lost. In our opinion, both approaches only lead to disguising the problem, 
but not to solving it. For this reason we address the overfitting in the network
topology itself. \\
\begin{table}[b!]
  \caption{BraTS18 evaluation for different segmentation approaches. Gaussian 
    Noise ($\sigma=0.02$) is added to training and validation data.}
  \label{tab:brats_noise_valtrain}
  \begin{center} 
    \begin{tabular}{l|c|c|c}
      \hline
      \noalign{\smallskip}
      Method & Enhancing & Complete  & Core\\
      \noalign{\smallskip}
      \hline
      \noalign{\smallskip}
      CMS   & $0.69$ & $0.82$ & $0.74$\\
      U-Net~ & $0.72$ & $0.87$ & $0.81$\\
      CascNN~ & $0.76$ & $0.89$ & $0.81$\\
      No NewNet~ & $0.75$ & $0.90$ & $0.84$\\
      \noalign{\smallskip}
      \hline
    \end{tabular}
  \end{center}
\end{table}

In the following we consider the No NewNet (without the adjustments suggested 
by the authors) as our baseline. Similar to our first experiments, we add 
Gaussian Noise with zero mean and standard deviation $\alpha=0.02$ and $\alpha=0.04$ to 
our validation data; see Tab.~\ref{tab:generalization}. It turns out that 
the model in its simplest form performs similar to our cascadic 
Mumford-Shah method when not much noise is present in the data. However, 
as soon as the noise is seriously altering the data distribution, 
the model prediction collapses and is outperformed by the classical approach.
Obviously, the generalization performance is limited and the network overfits
the training data.\\
\begin{table}[b!]
  \caption{BraTS18 evaluation for different adaptations of No NewNet. 
  Gaussian Noise is added to the validation data.}
  \label{tab:generalization}
  \begin{center} 
    \begin{tabular}{l|c|c|c}
      \hline
      \noalign{\smallskip}
      \multicolumn{4}{c}{Slight disturbance ($\sigma=0.02$)}\\
      \noalign{\smallskip}
      \hline
      \noalign{\smallskip}
      Method & Enhancing & Complete  & Core\\
      \noalign{\smallskip}
      \hline
      \noalign{\smallskip}
      CMS   & $0.69$ & $0.82$ & $0.74$\\
      Baseline~ & $0.69$ & $0.82$ & $0.76$\\
      Baseline + SWA & $0.73$ & $0.86$ & $0.81$\\
      Baseline + OctConv~ & $0.71$ & $0.84$ & $0.77$\\
      Baseline + OctConv + SWA~ & $0.74$ & $0.88$ & $0.83$\\
      Baseline + OctConv + SWA + post~ & $0.78$ & $0.89$ & $0.84$\\
      \noalign{\smallskip}
      \hline
      \noalign{\smallskip}
      \multicolumn{4}{c}{Moderate disturbance ($\sigma=0.04$)}\\
      \noalign{\smallskip}
      \hline
      \noalign{\smallskip}
      Method & Enhancing & Complete  & Core\\
      \noalign{\smallskip}
      \hline
      \noalign{\smallskip}
      CMS   & $0.67$ & $0.79$ & $0.71$\\
      Baseline~ & $0.66$ & $0.72$ & $0.70$\\
      Baseline + SWA & $0.70$ & $0.81$ & $0.76$\\
      Baseline + OctConv~ & $0.69$ & $0.77$ & $0.72$\\
      Baseline + OctConv + SWA~ & $0.71$ & $0.82$ & $0.79$\\
      Baseline + OctConv + SWA + post~ & $0.73$ & $0.85$ & $0.81$\\
      \noalign{\smallskip}
      \hline
    \end{tabular}
  \end{center}
\end{table}
\newline\noindent In a first step, we apply stochastic weight averaging (see 
Sec.~\ref{sec:swa}) with a cycle length of $10$ after $75\%$ of the training 
epochs. This adaptation of the training cycle obviously has an massive 
influence on the generalization behavior. The averaging of multiple minima in 
the loss surface allows the model to cope well with the disturbed data while 
neither the model capacity nor the training time is increased: While the model 
improves in the first scenario by $\approx 0.04$ on average, the 
performance gain of $\approx 0.06$ 
in the second setting with heavier noise disturbances is massive.\\
\noindent
Octave convolutions (see Sec.~\ref{sec:octconv}) have already shown in various 
applications that, in addition to a massive reduction in model size, 
they also contribute to improving generalization performance \cite{chen2019drop}. Consequently, we exchange all ordinary convolutions in the model by 3D octave convolutions ($\alpha=0.75$). Although this minor change does not alter the network topology, in both settings the performance increases 
by $\approx0.02$ and $0.03$ in Dice score over the baseline approach. This 
indicates a better generalization at inference to the validation data. The 
combination of stochastic weight averaging and the inclusion of frequency-aware 
octave convolutions leads to a improvement of $\approx 0.06$ with slight and 
$\approx 0.08$ moderate disturbances over the baseline.
\\
\noindent
The results of both of these modifications let us conclude that overfitting 
is indeed a serious problem - otherwise our changes would not lead to such 
drastic improvements.\\
Afterwards, we use the sparsified results as input for our semi-automatic
segmentation approach; see Sec.~\ref{sec:ec:post}. 
In the first setting, this postprocessing step mainly corrects for 
false positive labels of the enhancing tumor core; 
see Tab.~\ref{tab:generalization}. However, in the second scenario the robust
energy formulation stabilizes the segmentation and increases the overall 
performance for all classes.\\
In the end, we evaluated our final model (No NewNet+OctConv+SWA+post) on the 
original BraTS data without additional noise. We did not observe any drop in 
its performance: With Dice scores of $0.79$ for the enhancing tumor core, 
$0.90$ for the whole tumor and $0.85$ for the tumor core our approach is on par 
with current state-of-the-art approaches.\\
Please note, that the intention of this work is not to publish the next 
neural network trained on BraTS data. We rather want to highlight that 
generalization is a serious problem when improving on the benchmark metrics 
is the main goal. Of course one 
might argue, that those networks are never meant to be directly applied in a 
clinical setting. We only partly agree with this opinion. First, BraTS was 
originally designed to allow for a fair comparison and especially to push 
research in the direction of brain tumor segmentation. In this context, neural 
networks that can only be applied to benchmark data sets counteract the goal of a 
medical image segmentation challenge. Second, networks with a high performance 
on these data sets should at least perform similar on real data - but in our 
experiments, all approaches except the No NewNet architecture showed a much 
lower performance than in the benchmark setting - and even dropped below our 
method that exclusively rely on reliable prior information. Third, we are 
deeply convinced that increasing complex models do not lead to a satisfying 
real-world performance. Similar to Isensee et al. \cite{isensee2018nnu}, we 
implemented several suggested network extensions and found them mostly 
pointless. Our experiments even indicate, that they might be harmful as soon as 
training and validation data are not generated by the exactly same 
distribution. Hence, we fully agree that a well trained U-Net architecture is 
sufficient to solve this segmentation task.\\
All in all, we improved the generalization performance of the No NewNet 
architecture by straight forward adjustments in the model and the training 
procedure itself. \\
Although we neither changed its topology nor did we need to include the
noise distribution in our training data, we could robustify the network while
improving its generalization performance. Since this model is actually only a 
slightly modified version of the original U-Net, our suggested modifications 
also apply to similar structures.

\section{Conclusions}
With our paper we have addressed the general problem of model overfitting
of deep neural networks in brain tumor segmentation. Although the basic
assumption to learn a class distribution from the training data is very powerful,
it is also an Achilles heel when training and validation data slightly differ.
In a first step, we added noise to the validation data. Unfortunately,
our evaluations showed that such small variations lead to a massive drop
in network performance for two of the three best performing methods of 
BraTS~2018. 
Afterwards, we analyzed the behavior of networks when training and validation
data both are disturbed in the same way. It turned out, that this additional
information allows the network to cope with noisy data. However, since adding 
noise to the training data can have massive side-effects, we suggested
several straightforward modifications to be included in network designs.
Last but not least, we showed that these adjustments dramatically improve
the generalization performance. Although we did not include the disturbance in
the training data, we could reach with the same network topology nearly the 
same performance than without adding noise. This leads us to the conclusion 
that, in principle, all extensions of a well trained U-Net architecture for 
brain tumor segmentation not only fail to improve the result, but also worsen 
the generalization performance. \\
In our ongoing research, we plan to investigate further simplifications of deep
neural network models and the application of our findings to different benchmark
data sets. Furthermore, we target the problem of architectural overfit of network
topologies with a data dependent design.

\section*{Acknowledgments}
N. Graf has received partly funding from the European Union’s Seventh Framework 
Program for  research,  technological  development  and  demonstration  (grant  
agreement  No 600841, CHIC, Computational Horizons in Cancer).  The research of 
J. Weickert received funding from European Union’s Horizon 2020 research and 
innovation program (grant agreement No 741215, ERC Advanced Grant INCOVID).
The authors state no conflict of interest and have nothing to disclose.

\bibliographystyle{unsrt}  

\bibliography{references}

\end{document}